\documentclass[fleqn,usenatbib]{mnras}

\usepackage{newtxtext,newtxmath}

\usepackage[T1]{fontenc}

\DeclareRobustCommand{\VAN}[3]{#2}
\let\VANthebibliography\thebibliography
\def\thebibliography{\DeclareRobustCommand{\VAN}[3]{##3}\VANthebibliography}

\usepackage{graphicx,times}
\usepackage{natbib}
\usepackage{amsmath}
\usepackage{multirow}
\bibpunct{(}{)}{;}{a}{}{,}
\defcitealias{EE1995}{EE1995}
\defcitealias{VV2003}{VV2003}
\defcitealias{SL2013}{SL2013}

\title[New method for corotation radius determination]{A new, purely photometric method for determination of resonance locations in spiral galaxies}

\author[A. A. Marchuk et al.]{Alexander A. Marchuk,$^{1,2}$\thanks{E-mail: aamarchuk+astro@gmail.com}
Aleksandr V. Mosenkov,$^{3}$
Ilia V. Chugunov,$^{1,2}$
Valeria S. Kostiuk,$^{2}$
\newauthor
Maria N. Skryabina,$^{1,2}$
Vladimir P. Reshetnikov$^{1,2}$
\\
$^{1}$Central (Pulkovo) Astronomical Observatory, Russian Academy of Sciences, Pulkovskoye chaussee 65/1, St. Petersburg 196140, Russia\\
$^{2}$Saint Petersburg State University, Universitetskĳ pr. 28, St. Petersburg 198504, Russia\\
$^{3}$Department of Physics and Astronomy, N283 ESC, Brigham Young University, Provo, UT 84602, USA\\
}

\date{Accepted XXX. Received YYY; in original form ZZZ}

\pubyear{2023}

\begin{document}
\label{firstpage}
\pagerange{\pageref{firstpage}--\pageref{lastpage}}
\maketitle

\begin{abstract}
The knowledge of the positions of the corotation resonance in spiral arms is a key way to estimate their pattern speed, which is a fundamental parameter determining the galaxy dynamics. Various methods for its estimation have been developed, but they all demonstrate certain limitations and a lack of agreement with each other. Here, we present a new method for estimating the corotation radius. This method takes into account the shape of the profile across the arm and its width and, thus, only photometric data is needed. The significance of the method is that it can potentially be used for the farthest galaxies with measurable spiral arms. We apply it to a sample of local galaxies from Savchenko et al. and compare the obtained corotation radii with those previously measured in the literature by other methods. Our results are in good agreement with the literature. We also apply the new method to distant galaxies from the COSMOS field. For the first time, corotation locations for galaxies with photometric redshifts up to $z\sim0.9$ are measured.
\end{abstract}




\section{Introduction}           
\label{sect:intro}

\begin{figure*}
   \centering
   \includegraphics[width=1.95\columnwidth, angle=0]{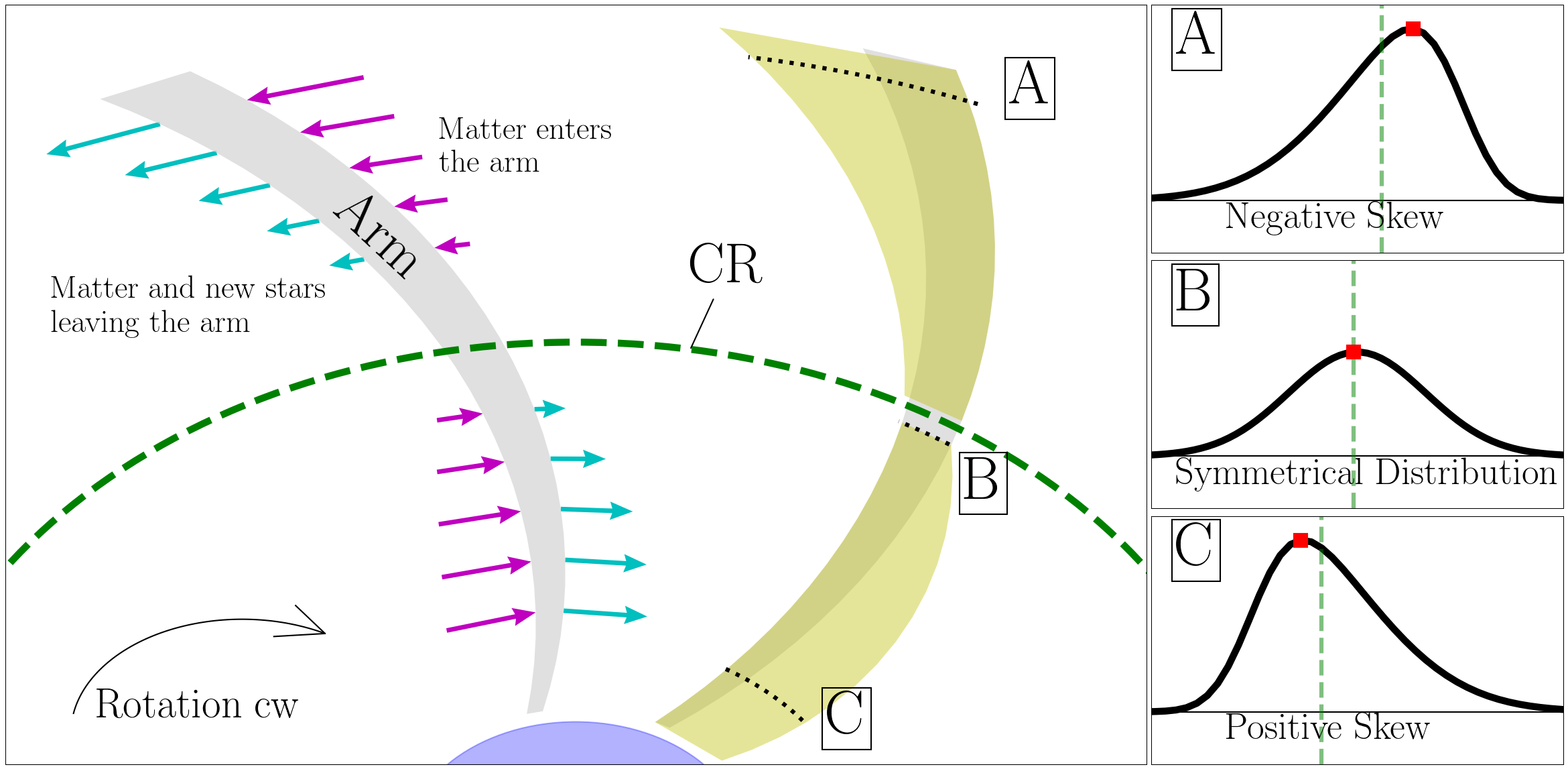}
   \caption{Schematic explanation of the method. The green dashed line represents the corotation circle, the bar/bulge/galactic center is schematically represented by a blue oval. The left arm is when the matter is not yet affected by the density wave, whereas the right arm is the same arm when the matter has been affected by the density wave.
   The magenta arrows indicate the entering directions of the matter into the arm, and the cyan arrows show the exiting directions. The length of the arrows corresponds to the relative velocity between the disc and the arms, with longer arrows representing higher velocities. Subplots on the righthand side show azimuthal profiles across the arm in the corresponding locations.
   }
   \label{fig:scheme}
\end{figure*}

The origin of spiral structure in disc galaxies is a long-standing issue which is still not fully understood. Nowadays, mostly two theories for its explanation are discussed in the literature. The first of them is based on the well-known theory of quasi-stationary density waves~\citep{Lin&Shu,BertinLin1996}, in which spiral arms are considered density waves. Probably, the best illustrative presentation of this theory was given in \citet{Kalnajs73}. The angular rotation velocity of spiral pattern $\Omega_\mathrm{p}$, that is the frequency of precession of the stellar orbits for the spiral arms, is assumed to be constant at any distance from the centre of the galaxy. The theory of transient spirals, also called recurrent or dynamic, assumes that the spiral structure is dynamic in nature, and its angular speed changes mostly differentially with the one for the disc \citep{Baba13,Sellwood10}. In this case, the spiral arms do not visually appear continuous, but rather consist of individual segments. 
\par
Despite the strength of arguments in theoretical calculations, it is a rather difficult task to confirm the aforementioned theories using observations. A long-lived spiral density wave can only be obtained with swing amplification \citep{Toomre81}, which takes into account the effects that amplify the wave moving in the radial direction at resonances. One of such resonances is located at the distance where the angular velocities of the spiral arms and the disc coincide. It is called the corotation radius (CR hereafter) or corotation resonance. Finding reliably determined corotation radii and the corresponding angular velocities of spiral pattern is of great importance for the following reasons. First, the existence or absence of a localized CR is one of the few observational confirmations in favour of one or another theory of the origin of spiral pattern. Secondly, the presence of a CR may influence the chemical evolution in the disc since this radius is thought to possibly separate the disk into two isolated regions~\citep{Vila-Costas92,SL2013}. Also, the position of the CR is associated with the properties of the stellar orbits and their stability~\citep{Contopoulos13}. Finally, the corotation resonance is associated with the transfer of angular momentum in the disc, a rather important process for understanding its secular evolution \citep{SellwoodBinney02}. There are many other relevant problems in astrophysics where the estimation of the CR (or, equally, $\Omega_\mathrm{p}$) is vital, e.g. determination of the stability in spiral arms \citep{Inoue18} or in recent observations of the star formation in NGC~628 carried by the JWST \citep{Spurring}.
These examples show the exceptional importance of studying corotation resonances and the need for their reliable determination in galaxies. 
\par
There are many methods for measuring the position of CR, such as the Tremaine-Weinberg method \citep{1984ApJ...282L...5T,2009ApJ...702..277M,cuomo2020,2021AJ....161..185W} based on finding the age gradient of stars across the arm \citep{Puerari97,Tamburro08,Egusa09,2015MNRAS.450.1799S}, measuring potential-density phase-shift \citep{2007AJ....133.2584Z,2009ApJS..182..559B}, probing morphological features near resonances \citep{EE1995} tracing velocity sign reversal \citep{2011ApJ...741L..14F,2014ApJS..210....2F}, and others. At the same time, the results obtained with their aid are contradictory \citep{Vallee20} and unreliable \citep{2021AJ....161..185W,2023MNRAS.524.3437B}. For this reason, new methods for finding the CR are in demand and their number continues to increase (see, for example, \citealt{Pfenniger23}). In this paper, we propose a new independent method for measuring the CR in spiral galaxies using information about the variation of the half-widths of the spiral arms with radius. The advantage of our method is that it only utilizes photometric data making it extremely useful for studying resolved galaxies with measurable spiral arms at high redshift.
\par

\section{The method}
\label{sect:method}

The main idea of the proposed method is illustrated in Fig.~\ref{fig:scheme}. As in many other methods, it utilizes the fact that the spiral arms in a spiral galaxy rotate faster than the disc outside the CR and slower than the disc in the inner region inside the corotation circle. Since the spiral pattern maintains a constant angular speed $\Omega_\mathrm{p}$ with radius, the relative velocity between the disc and the arms increases with greater distance from the CR. Thus, the amount of matter, entering the arm body, upstreams before the CR and downstreams outside the CR, also increasing with distance. When the galactic gas enters the density wave potential minimum, it undergoes compression and forms new stars in the number dependent on the injected matter \citep{2021A&A...656A.133Q}. After the formation, in most cases a newborn star of intermediate or small mass leaves the arm following the rotation of the disc. These facts are schematically illustrated in Fig.\ref{fig:scheme}, left part. 
\par
During further rotation, these stars form a ``trail'' observed on the front side in the inner part of the arm and on the rear side in the outer part of the arm, as shown in Fig.~\ref{fig:scheme}, left part and right arm. Thus, for stars that live long enough and with the condition of constant $\Omega_\mathrm{p}$, we can formulate an observational hypothesis about azimuthal profiles across the spiral arm. Along an arm, rotating in clockwise direction as in Fig.~\ref{fig:scheme}, we should observe a negatively skewed profile within the corotation circle and a relatively symmetrical one closer to the CR. In other words, the inward width of the arm should be smaller than the outward. After the CR, the arm profile should become more and more positively skewed, with a longer tail and the steeper opposite side, since the size of the effect depends on the exiting velocity. Such profiles are schematically illustrated in the right part of Fig.~\ref{fig:scheme}. Note that similar considerations for one $N$-body model were presented in \cite{Debattista2014}.
\par
From the practical side, the realization of the method is as follows. First, we need to measure azimuthal profiles across individual arms. There are plenty of ways to do that, mentioned in the next section. After that, the degree of skewness for each profile should be measured. In principle, the parameters $h_3$ (the skewness for measuring asymmetric deviations) and $h_4$ (the kurtosis for estimating symmetric deviations) can be used \citep{1993ApJ...407..525V}, but they may be difficult to interpret. Instead, for each profile, we estimate its peak's position and where its median is located, then plot them in circular coordinates. Assuming the trailing nature of the spiral arm and taking into account the direction of rotation, the intersection of both curves should mark the CR position. 


\begin{figure}
   \centering
   \includegraphics[width=0.95\columnwidth, angle=0]{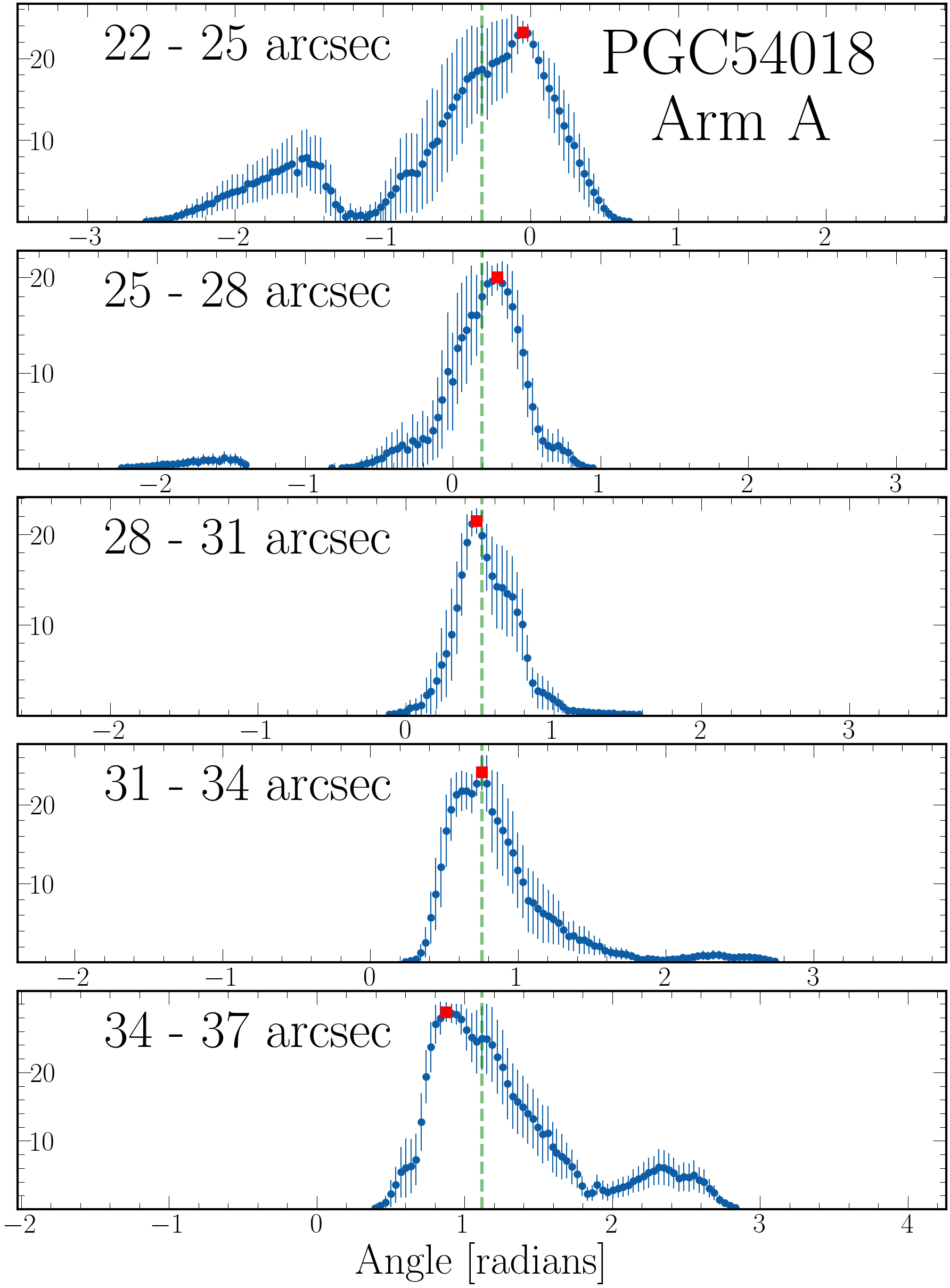}
   \caption{An example of the method application for PGC\,54018 rotated in the clockwise direction. Azimuthal profiles and their peaks (red) and medians (green) are depicted. The CR is measured to be between 26 and 33~arcsec.} 
   \label{fig:example}
\end{figure}

\section{Validation of the method and its application}
\label{sec:validation}

To apply the proposed method, we need to have azimuthal profiles of a spiral arm within its width across the galaxy. The widths of the spiral arms in galaxies are mostly unknown and are rarely studied in general. Here, we can mention   \cite{1982ApJ...253..101K}, \cite{2015ApJ...800...53H} and \cite{2022MNRAS.514L..22S}. The reason for such a lack of attention to this topic is that the spiral arms contain a lot of dust and enhanced star formation, and, thus, their photometric profiles are noisy and difficult to measure. 
As an example of such noisy profiles, see figure~7 in \cite{2011MNRAS.414..538K}. 
Often used for similar tasks, Fourier transform is also of little help here because 1) spiral arms can have complex non-symmetrical profiles and 2) such a transform is often difficult to apply (see, for example, results of \citealt{2018ApJ...862...13Y}). The novelty of the approach used in \cite{savchenko} is that an arm is consequently dissected with several parallel cuts, carefully avoiding and masking bright areas. The obtained slices are then used for averaging and profile fitting with an analytical function. We refer the reader to \cite{savchenko} and \cite{2020RAA....20..120M} for details of the method. 
\par
\par
The advantage of the \cite{savchenko} technique is that if the slices are sufficiently densely distributed and the arm is not extensively contaminated by dust, then we, in fact, have a good model of the arm where peculiarities are smoothed out. This data can then be used for applying our method. For doing so, we employ a spline interpolation between the profiles in individual slices, constructing a model of the individual arm. We then divide it into annuli and create azimuthal profiles, shown for the example galaxy PGC\,54018 in Fig.~\ref{fig:example}. Note the similarity between the observed profiles for PGC\,54018 and the idealized expectations presented in Fig.~\ref{fig:scheme}. 
\par
\cite{savchenko} applied their method to 155 face-on galaxies in three optical bands using SDSS data \citep{2000AJ....120.1579Y}. It is important to note that optical bands are convenient for the proposed method since the light in these bands is emitted by a mixed stellar population, including the old stars. A noticeable number of those galaxies have CR measured in the literature using other methods. To demonstrate the validity of our new method, we apply it to galaxies from \citep{2000AJ....120.1579Y} and compare our results with those from the literature. 
\par
We apply our method to images in the $r$ band for objects satisfying several conditions. First, the slices should be dense enough; otherwise, the arm's model will contain ``gaps'' in its body. Second, only galaxies where all traced arms have CR (and these measurements agree with each other) are selected. Finally, the predicted CR value should be stable. To do so, we try different annulus sizes and estimate the CR measurement error using a common Monte-Carlo simulation, where points in each profile (such as in Fig.~\ref{fig:example}) in each realization are rearranged within their error margins. We assume the spiral arms to be trailing by nature.
\par
\begin{figure}
   \centering
   \includegraphics[width=0.95\columnwidth, angle=0]{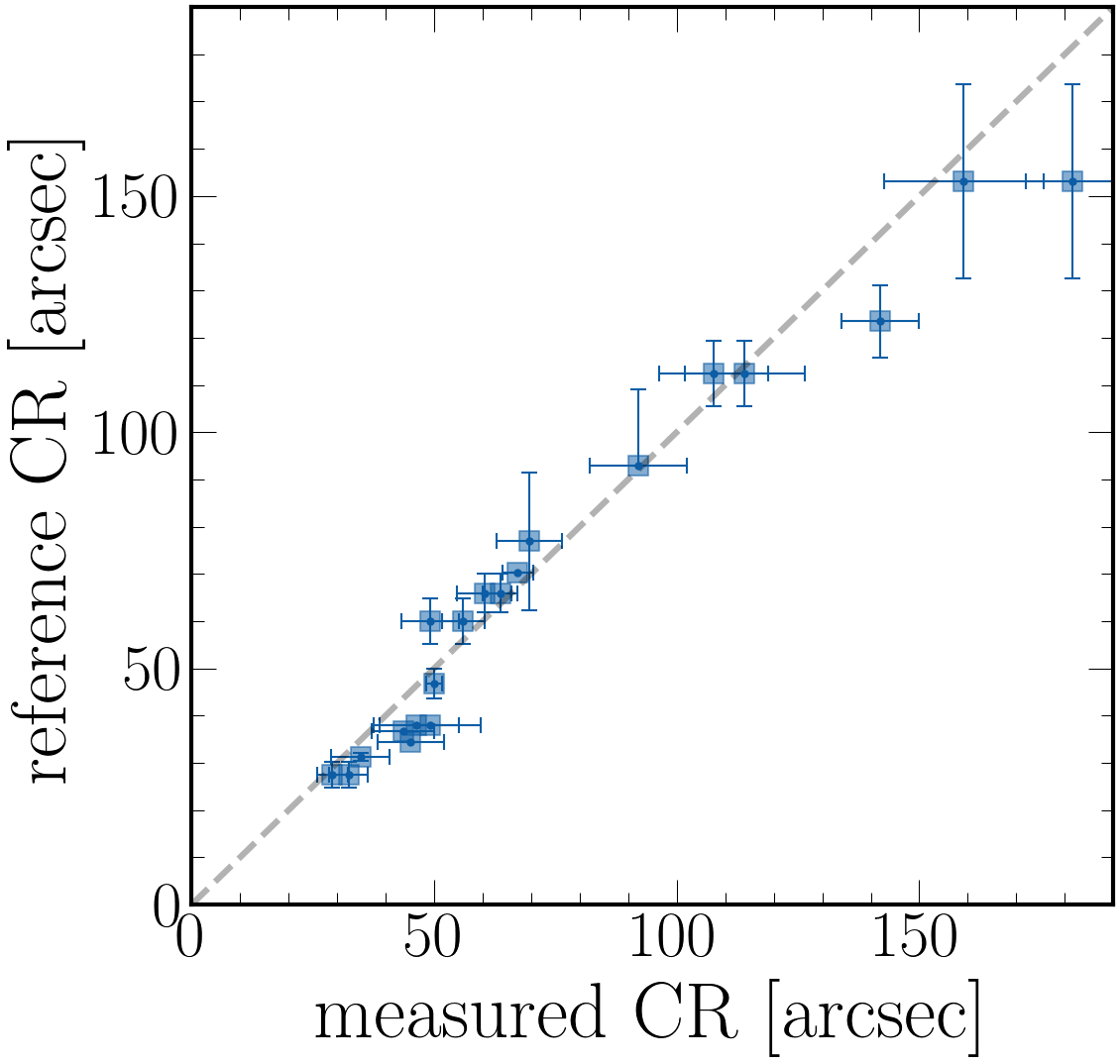}
   \caption{Agreement of the CR obtained by our method and reference values from the literature listed in Table~\ref{tab:valid}. Each point is given for one measured arm. The reference CR have been averaged if more than one CR is given in the reference.} 
   \label{fig:agreement}
\end{figure}

The results of our analysis for 13 galaxies that fulfill the mentioned conditions and with known previously CR data are presented in Table~\ref{tab:valid}. For individual arms of these galaxies, we list the CR obtained using our method and one or several reference values and their errors from the literature. Note, that the CR in half of the galaxies is confirmed with more than one measurement from the literature, increasing the reliability of the result. The measurements for individual arms agree with each other, with the only exception being PGC\,4946 where the measured parts of the arms do not intersect much with radius. In Fig.~\ref{fig:agreement}, we display the agreement of the results of our method and those from the literature noticing that points lie close to the diagonal line. The good consistency of the results allows us to conclude that the proposed method is valid, robust and, most importantly, is easy to apply.
\par
Before we proceed, several important notes should be made. For the galaxy PGC\,54445, we see signs of CR to be located near $\approx96$~arcsec, which agrees with the value from \cite{EE1995}, but the signal is rather weak and not confirmed by the Monte-Carlo simulation. We also see several galaxies with a strong signal, which is consistent among the individual arms and with the literature, but, at the same time, demonstrate profiles reversed to what is presented in Fig.~\ref{fig:scheme}. In other words, in these galaxies, the arms may be leading instead of trailing. An example of such a galaxy is PGC\,2949, where the arms appear to be rotating clockwise (in all cases where north is up and east is to the left), but the locations of the peaks and the medians suggest the opposite. Its CR value equals approximately 40~arcsec that is confirmed by \cite{cuomo2020}. We do not count such ``suspicious'' galaxies as supporting evidence for our method but acknowledge that CR can, in principle, still be detected with our method and leave this for future research.

\begin{table}
	\centering
	\caption[]{CR for validation galaxies from \cite{savchenko}. 
 The abbreviations EE1995, VV2003 and SL2013 denote \cite{EE1995}, \cite{VV2003} and \cite{SL2013}, accordingly. All the values are given in arcseconds; the reference CR values without errors are those from \cite{2009ApJS..182..559B} where error bars were not estimated.\label{tab:valid}}
	\begin{tabular}{lllll} 
		\hline
		PGC &  Arm & $\mathrm{CR}$ & $\mathrm{CR_{ref}}$ & Reference\\
            \noalign{\smallskip}
		\hline
  \noalign{\smallskip}
		2901 & A & $60.4\pm5.6$ &$65.9^{+4.1}_{-4.1}$ & \cite{2014ApJS..210....2F}\\
  &  B & $63.6\pm3.6$ &   &  \\
\hline
4946 & A & $67.2\pm3.2$ &$70.3^{+0.0}_{-0.0}$ & \cite{2009ApJS..182..559B}\\
  &  B & $45.2\pm6.8$ &$34.5^{+0.0}_{-0.0}$ & \cite{2009ApJS..182..559B}\\
\hline
8974 & A & $49.2\pm6.0$ &$60.0^{+4.8}_{-4.8}$ & \citetalias{EE1995} \\
  &  B & $56.0\pm4.4$ &  &   \\
\hline
23028 & A & $50.0\pm1.6$ &$47.4^{+3.2}_{-3.2}$ & \cite{2014ApJS..210....2F}\\
  &  &  & $46.0^{+0.0}_{-0.0}$ & \citetalias{VV2003}\\
\hline
27077 & A & $181.6\pm9.6$ &$158.2^{+25.3}_{-25.3}$ & \citetalias{SL2013}\\
  &  &   &$153.7^{+20.2}_{-20.2}$ & \cite{2020MNRAS.496.1610A}\\
  &  &   &$153.0^{+16.8}_{-16.8}$ & \citetalias{EE1995} \\
  &  B & $159.2\pm16.4$ &$158.2^{+25.3}_{-25.3}$ & \citetalias{SL2013}\\
  &  &   &$153.7^{+20.2}_{-20.2}$ & \cite{2020MNRAS.496.1610A}\\
  & &  &$153.0^{+16.8}_{-16.8}$ & \citetalias{EE1995}
  \\
\hline
27777 & A & $43.6\pm6.4$ &$36.7^{+0.0}_{-0.0}$ & \cite{2009ApJS..182..559B}\\
\hline
31968 & A & $142.0\pm8.0$ &$127.3^{+2.3}_{-2.3}$ & \cite{2011ApJ...741L..14F}\\
  &  &  
 &$119.6^{+12.9}_{-12.9}$ & \cite{2009ApJ...702..277M}\\
\hline
34232  & A & $34.8\pm6.0$ &$32.1^{+1.8}_{-1.8}$ & \cite{2008MNRAS.388.1803R}\\
  &  &   &$30.5^{+0.0}_{-0.0}$ & \cite{2009ApJS..182..559B}\\
\hline
42168 & A & $69.6\pm6.8$ &$71.1^{+8.4}_{-8.4}$ & \cite{2008MNRAS.388.1803R}\\
  &  &   &$80.7^{+17.5}_{-17.5}$ & \cite{2021AJ....161..185W}\\
  &  &   &$81.3^{+18.4}_{-18.4}$ & \cite{2020MNRAS.496.1610A}\\
\hline
49514 & A & $92.0\pm10.0$ &$93.0^{+0.0}_{-16.0}$ & \cite{2020MNRAS.496.1610A}\\
\hline
54018 & A & $29.0\pm3.2$ &$27.5^{+2.8}_{-2.8}$ & \cite{2014ApJS..210....2F}\\
  &  B & $32.4\pm4.0$ &   & \\
\hline
60459 & A & $114.0\pm12.4$ &$131.7^{+0.0}_{-0.0}$ & \cite{2009ApJS..182..559B}\\
  &   &   &$93.0^{+13.8}_{-13.8}$ & \citetalias{EE1995}\\
  &   B & $107.6\pm11.2$ &$93.0^{+13.8}_{-13.8}$ & \citetalias{EE1995}\\
  &  &   &$131.7^{+0.0}_{-0.0}$ & \cite{2009ApJS..182..559B}\\
\hline
72263 & A & $46.4\pm8.8$ &$38.0^{+0.0}_{-0.0}$ & \cite{1998AJ....116.2136A}\\
 &  B & $49.2\pm10.4$ &  & \\
		\hline
	\end{tabular}
\end{table}

The sample in \cite{savchenko} is not the only one to which the adopted slicing technique has been applied. \cite{reshetnikov} processed the spiral arms of distant galaxies in the HST COSMOS field \citep{2007ApJS..172..196K}. The final sample contains 102 galaxies with a distinct two-armed spiral pattern, and the image scale is 0.03~arcsec/pix. All photometric redshifts were taken from the COSMOS2020 catalog \citep{2022ApJS..258...11W} and have an estimated accuracy of around 1\%. The median redshift of these galaxies is $z\sim 0.37$, which, taken into account with the wavelength of the F814W filter, is suitable for an analysis with the aid of our method. In fact, these galaxies appear much smoother than those in the local Universe, and, thus, our new method is much easier to apply than to nearby objects. 
\par
Using our method, identified a CR position in 64 out of 102 from \cite{reshetnikov}, i.e. in most of the objects. The results for individual arms were averaged, and the spirals are assumed to be trailing. As a simple sanity check, we observe a tendency of the CR (expressed in arcseconds) to decrease with higher $z$, which is expected. The number of galaxies with clockwise and counterclockwise directions of rotation is almost equal. Although the CR in all these galaxies is estimated in this paper for the first time, for some galaxies we find supporting evidence for the location of a resonance based on the visual inspection of the arms. For example, the galaxy with id\,1377853 demonstrates a significant change in the arms' appearance at the CR, as do several other galaxies. We present the results of measuring the CR for the 64 galaxies in the online material.

\section{Discussion and conclusions}
\label{sect:conclusions}

For galaxies from the \cite{reshetnikov} sample, we add the effective radius $R_\mathrm{eff}$ from the catalogue in \cite{2022ApJS..258...11W}, which is the radius enclosing 50\% of the total flux in the PSF-homogenized $K_s$-band image. The CR mostly lies not far from the $R_\mathrm{eff}$, and their ratio for 50\% of galaxies in the interquartile range lies within $R_\mathrm{CR}/R_\mathrm{eff} = 0.95 - 1.35$. \cite{EE1995} identified a CR at $\simeq 0.5\times R_{25}$ (here $R_{25}$ is the so-called optical radius measured at a surface brightness level of 25~mag\,arcsec$^{-2}$ in the $B$-band) in a significant number of galaxies out of 173 objects. Since, on average, $R_{25}\simeq4-5$~disc scale lengths \citep{2011ARA&A..49..301V} and if we assume $R_\mathrm{CR}/R_\mathrm{eff} \approx 1.2$ and $R_\mathrm{eff}\approx 1.7$~disc scale lengths for galaxies with a small bulge, we then derive $R_\mathrm{CR}\simeq 0.5\times R_{25}$, which broadly agrees with \cite{EE1995} findings (see also \citealp{Tamburro08}). We also identify a weak positive correlation between the CR and $R_\mathrm{eff}$, which suggests that the central mass concentration may be connected with the spiral arms \citep[see e.g.][]{2005MNRAS.359.1065S,2006ApJ...645.1012S,2014ApJ...795...90S}, but whether this is true or not is a matter of discussion \citep{2017MNRAS.472.2263H}.
\par
The proposed method has its drawbacks. Firstly, a galaxy can demonstrate more than one CR, and not only in different morphological features i.e. in bar and in spirals, but in individual parts of spiral pattern as well \citep[e.g.][]{2014ApJS..210....2F,2009ApJ...702..277M}. It is unclear how these parts of spiral pattern connect in such a case and how this situation affects the measured widths. Second, the rotation curve may have a complex shape and, hence, its angular speed $\Omega(r)$ may demonstrate ``bumps'', which would complicate the profiles presented in Fig.~\ref{fig:scheme}. However, usually $\Omega(r)$ demonstrates a monotonic decrease, so this should not be an issue. We also suppose non-circular motions in the disc to be negligible. We do not know, in general, how prominent the inner and outer edges of the arm should be and what processes govern them (see, for example, discussion in \citealt{2019ApJ...874..177M}). Next, the technique of \cite{savchenko}, while can be automated to some extent, still needs a lot of manual work and is, thus, relatively hard to apply to a large sample. Stars may change the direction of their motion and radially migrate inside the spiral arms, and we do not know the prominence of this effect. Finally, we do not know how the uncertainties of the orientation parameters, instrumental properties (such as the PSF and angular resolution), correction for distance, or various interpolation techniques in model building would influence the obtained CR value and its uncertainties. These are significant questions for future tests and studies. However, as demonstrated by Fig~\ref{fig:agreement}, the good agreement with the literature suggests that the method works correctly and gives reliable results.
\par
Despite the fact that the density waves theory was developed more than half a century ago, the key aspects of the process of how the spiral arms develop and retain their nature are still missing. Since most of the galaxies in the local Universe contain spirals of a different kind, and the JWST is able to detect them at least up to $z \approx 3$ \citep{2023ApJ...942L...1W}, the question under consideration is of great importance. Knowing the corotation radius is an essential aspect for solving the problem of the spiral structure formation, along with other various questions related to secular evolution in the discs \citep{SL2013,Contopoulos13,SellwoodBinney02}. 
That is why tens of various methods for the CR estimation have been developed over the years \citep{1984ApJ...282L...5T,EE1995,2011ApJ...741L..14F,SL2013,Puerari97} and information about the CR in hundreds of galaxies have been collected. However, these measurements often contradict each other due to the unknown validity of the assumptions used and a generally very small size of the measured effect.
\par
In this short paper, we have presented a new method for determination of the corotation radius (CR). The advantage of the new method is that it is based on the measurements of the azimuthal widths across the spiral arms. This allows us to measure the CR even in very distant galaxies. The idea of the method is simple and based on the fact that the difference in angular velocity of the spiral arms and the disc increases with greater distance from the CR. Therefore, this determines in which direction and how fast the stars leave the arm forming a ``trail''. This feature is visible in the azimuthal profiles in the form of skewness in different directions before and after the corotation, which can be easily measured. We apply the proposed method to the spiral arms measured previously in \cite{savchenko} with the slicing technique and compare the measured CR with those compiled from the literature. The concordance between the results, illustrated in Fig.~\ref{fig:agreement}, unequivocally states that the method works correctly and is able to determine the CR. 
\par
In the final part of the work, we have applied the proposed method to distant spiral galaxies identified in the COSMOS field by \cite{reshetnikov}. 
Using our method, we have identified the CR position in more than half of them which demonstrates the applicability of the method to study very distant spiral galaxies. 
\par
This paper mostly serves as a proof-of-concept which presents the main idea of the method and its general reliability. In the future, we are about to test the method's limitations, which will potentially increase the number of galaxies with measured CR. In our further studies, we will compare the retrieved parameters of the spiral pattern with the general galactic properties and attempt to link them with the different theories of the spiral structure generation.

\section*{Acknowledgements}

We thank the anonymous referee for the review and highly appreciate the comments and suggestions. This work was supported by the Russian Science Foundation (project no. 22-22-00483).

\section*{Data Availability}
 
Table results for \cite{reshetnikov} sample is available in the online material. The data underlying this article will be shared on reasonable request to the corresponding author.



\bibliographystyle{mnras}








\bsp	
\label{lastpage}
\end{document}